\documentclass[aps,pre, reprint,superscriptaddress]{revtex4-1}
\usepackage{amsmath}
\usepackage{mathtools, cuted}
\usepackage{lipsum}
\usepackage{lineno,hyperref}
\usepackage{cuted}
\usepackage[utf8]{inputenc}
\usepackage{verbatim}
\begin{document}

\title{Neutrino Oscillations in a Quantum Processor}

\author{C.A. Arg{\"u}elles}
\affiliation{Massachusetts Institute of Technology, Cambridge, MA 02139, USA}

\author{B.J.P. Jones}
\affiliation{Department of Physics, University of Texas at Arlington, Arlington, TX 76019, USA}

\begin{abstract}
Quantum computing technologies promise to revolutionize calculations in many areas of physics, chemistry, and data science.  Their power is expected to be especially pronounced for problems where direct analogs of a quantum system under study can be encoded coherently within a quantum computer.  A first step toward harnessing this power is to express the building blocks of known physical systems within the language of quantum gates and circuits. In this paper, we present a quantum calculation of an archetypal quantum system: neutrino oscillations. We define gate arrangements that implement the neutral lepton mixing operation and neutrino time evolution in two-, three-, and four-flavor systems. We then calculate oscillation probabilities by coherently preparing quantum states within the processor, time evolving them unitarily, and performing measurements in the flavor basis, with close analogy to the physical processes realized in neutrino oscillation experiments. Evaluations on publicly available quantum processors obtain excellent agreement with classical calculations.  We provide recipes for modeling oscillation in the standard three-flavor paradigm as well as beyond-standard-model (BSM) scenarios, including systems with sterile neutrinos, non-standard interactions, Lorentz symmetry violation, and anomalous decoherence.
\end{abstract}

\maketitle

\section{Introduction}

The unexpected and Nobel Prize-winning discovery of neutrino oscillations~\cite{Kajita:2016cak} has led to a program of experiment and theory that has shaped understanding of the role of neutrinos in the Universe. The spontaneous transition of neutrino flavor over macroscopic distances, a phenomenon known as neutrino oscillations due to its periodic behavior, demonstrates that neutrinos have masses that are non-zero but uniquely small. This smallness suggests connections to high-scale physics~\cite{gell1979ramond,yanagida1979proceedings,mohapatra1981neutrino}, and may be related directly to the predominance of matter over antimatter abundances in the Universe~\cite{Fukugita:1986hr}. Studies of neutrino oscillations have thus contributed and will continue to contribute greatly to our understanding of nature.

Experiments measure neutrino oscillations by studying a neutrino beam's flavor composition at different energies, $E$, and baselines, $L$~\cite{Nakamura:2010zzi}.
Oscillation refers to spontaneous transformation
between the three neutrino flavors -- $\nu_{e}$, $\nu_{\mu}$, and $\nu_{\tau}$ -- 
during flight. This is due to de-phasing of the neutrino wave functions during propagation arising from a misalignment between the flavor and mass bases. 
In the absence of strong matter
interactions, and when two neutrino mass states dominate the oscillation, a sinusoidal flavor variation as a function of $L/E$ is characteristic. For oscillations in matter~\cite{Wolfenstein:1977ue,Mikheev:1986gs} and with three neutrinos participating, more complex functional forms are observed~\cite{giunti2007fundamentals}. 
Well-studied neutrino sources, in which neutrino flavor changing has been observed, include $\nu_{\mu}$ and $\bar{\nu}_{\mu}$ production by decays of
charged pions from accelerators~\cite{Adamson:2011qu,Adamson:2016tbq} or in cosmic-ray air showers~\cite{Fukuda:1998mi,TheIceCube:2016oqi},
production of $\bar{\nu}_{e}$ by fission in nuclear reactors~\cite{Eguchi:2002dm,An:2012eh},
and of $\nu_{e}$ by nuclear fusion in the Sun~\cite{Davis:1968cp}.
Neutrino oscillations have been shown to violate the Leggett-Gaarg inequality~\cite{Formaggio:2016cuh}, a time domain version of Bell's classic argument~\cite{bell2001problem}, which illustrates that they are a truly quantum mechanical phenomenon with no possible description in terms of hidden classical variables. 

For neutrino oscillations to be observable, quantum coherence
between the neutrino mass basis states must be maintained over the flight distance of the neutrino~\cite{Akhmedov:2009rb}, which in some experiments is thousands of kilometers.
Neutrinos are thus very-long-baseline quantum interferometers, and have they been used as such to perform fundamental tests of quantum mechanics~\cite{Lisi:2000zt,Fogli:2003th} and Lorentz invariance~\cite{Kostelecky:2003cr,Ellis:2008fc,Aartsen:2017ibm}, in order to search for evidence of quantum gravity~\cite{Mavromatos:2004sz,Alfaro:1999wd} and violations of the equivalence principle~\cite{Gasperini:1988zf}.
The expected decoherence of oscillating neutrinos via wave-packet separation has been studied theoretically~\cite{Jones:2014sfa}, but not yet observed in experiments. 

Although quantum computing has been predominantly associated with extensive database queries~\cite{Grover:1996rk} or number factorization~\cite{Shor:1994jg}, it was realized early in its conceptual development that a natural connection between simulation of quantum systems and quantum computers exists~\cite{Feynman:1981tf}; see~\cite{Georgescu:2013oza} for a recent review on quantum simulation.
An advantage of quantum computers over classical systems is the ability to perform actual Hamiltonian evolution, rather than emulate it.
In particle physics we often deal with high-particle multiplicity processes, {\it e.g.} in high-energy collider experiments quantum computers have been pointed to be advantageous in the simulation~\cite{Bauer:2019qxa} and reconstruction~\cite{Wei:2019rqy} of hadronic showers. 
Other examples of physics frontiers that may be substantially advanced by quantum computation include modeling in nuclear physics~\cite{Cloet:2019wre}, many-body effects in condensed matter systems~\cite{buluta2009quantum}, and quantum chromodynamics~\cite{wiese2014towards}, among others.

This work demonstrates for the first time the processing of three-neutrino flavor information in a quantum simulation, executing an analogous Hamiltonian evolution to generate neutrino flavor oscillations. Such encoding and evolution is a vital building block on which more advanced quantum simulations involving neutrino flavor can be constructed. Systems that could particularly benefit from quantum algorithmic approach to neutrino flavor evolution include those where collective neutrino oscillations~\cite{Sigl:1992fn,Duan:2010bg} are relevant, such as in Supernovae or the early Universe~\cite{Biondini:2017rpb}.
In these cases the quantum Boltzmann equations that yield the evolution of the neutrino population can only be solved approximately~\cite{Blanchet:2011xq} or by specialized numerical techniques~\cite{Hernandez:2016kel,Delgado:2014kpa}.  This work demonstrating encoding of neutrino flavor structure and oscillation represents an important first step toward addressing such problems using quantum processors.

Recently, publicly accessible quantum processors were made available online as part of the {\tt IBM-Q} project, and these can be used for novel research into quantum processing~\cite{Lesovik:2019klx}. 
Although the technology remains imperfect, with error rates per gate operation of ${\cal O}(0.1\%)$ and per qubit read of ${\cal O}(5\%)$ prohibiting very lengthy calculations, 
the platform provides a test bed for exploring quantum solutions to computational problems, and finding ways to re-express calculations in the language of quantum circuits.

In this paper, we present a quantum simulation of neutrino flavor oscillations. After illustrating how to encode the two-neutrino system evolution in a quantum computer with a single qubit, we proceed to implement the less intuitive three-neutrino system realized on a subspace of a two-qubit Hilbert space. The primary challenges involved are the implementation of the Pontecorvo–Maki–Nakagawa–Sakata (PMNS)~\cite{Nakamura:2010zzi} operation, which relates the flavor and mass neutrino eigenstates, and the time-evolution operator of the system in the computational basis.  After testing that our quantum circuit reproduces the quantum neutrino oscillation probability on the IBM-Q public quantum computer, we conclude with a brief discussion of how to include more complex phenomena including sterile neutrinos, matter effects, non-standard interactions, Lorentz symmetry violation, and decoherence within the quantum algorithm. 

\section{Two-flavor neutrino oscillation}

Two-flavor neutrino oscillations involve a Hilbert space of two dimensions. This can be represented on a single qubit, via the basis choice:
\begin{equation}
|0\rangle=|\nu_{1}\rangle=\left(\begin{array}{c}
1\\
0
\end{array}\right)\quad~{\rm and}~\quad|1\rangle=|\nu_{2}\rangle=\left(\begin{array}{c}
0\\
1
\end{array}\right).
\end{equation}
The rotation into the flavor basis requires a unitary operation via
the two-dimensional PMNS matrix. The reduced PMNS operation is defined such that $|\nu_{e}\rangle=U_{PMNS}^{2x2\dagger}|0\rangle$ and $|\nu_{\mu}\rangle=U_{PMNS}^{2x2\dagger}|1\rangle$. The most general unitary transformation applicable to a single qubit
system, which must be able to support the 2$\times$2 PMNS operation, is encoded in the IBM quantum computer by the three-parameter
$U3$ gate:
\begin{equation}
U3(\Theta,\phi,\lambda)=\left(\begin{array}{cc}
\cos\frac{\Theta}{2} & -\sin\frac{\Theta}{2}e^{i\lambda}\\
\sin\frac{\Theta}{2}e^{i\phi} & \cos\frac{\Theta}{2}e^{i\left(\lambda+\phi\right)}
\end{array}\right).
\end{equation}
For the two-neutrino system, oscillation probabilities depend only
on one of the parameters of $U3$, for
the following reasons~\cite{giunti2007fundamentals}:
\begin{enumerate}
\item The parameter $\phi$ can be removed by a re-defininition
of the $|\nu_{\mu}\rangle$ basis state via $|\nu_{\mu}\rangle\rightarrow e^{-i\phi}|\nu_{\mu}\rangle$.
This corresponds to re-phasing the charged
muon field, under which the Standard Model Lagrangian is invariant. Without loss of generality, we can set $\phi=0$.
\item The parameter $\lambda$ could similarly be removed by re-phasing the $|\nu_{2}\rangle$ field, $|\nu_{2}\rangle\rightarrow e^{i\lambda}|\nu_{2}\rangle$.
The Lagrangian is only invariant under this re-definition if the neutrinos are Dirac particles.  If they are
Majorana particles, on the other hand, this phase is physical and must be maintained in the Lagrangian. However, it can be shown the ``Majorana phase'' $\lambda$ does not influence neutrino oscillations~\cite{Giunti:2010ec} and general oscillation probabilities
can be calculated under the assumption $\lambda=0$.
\end{enumerate}

\begin{figure*}
\begin{centering}
\includegraphics[width=1.7\columnwidth]{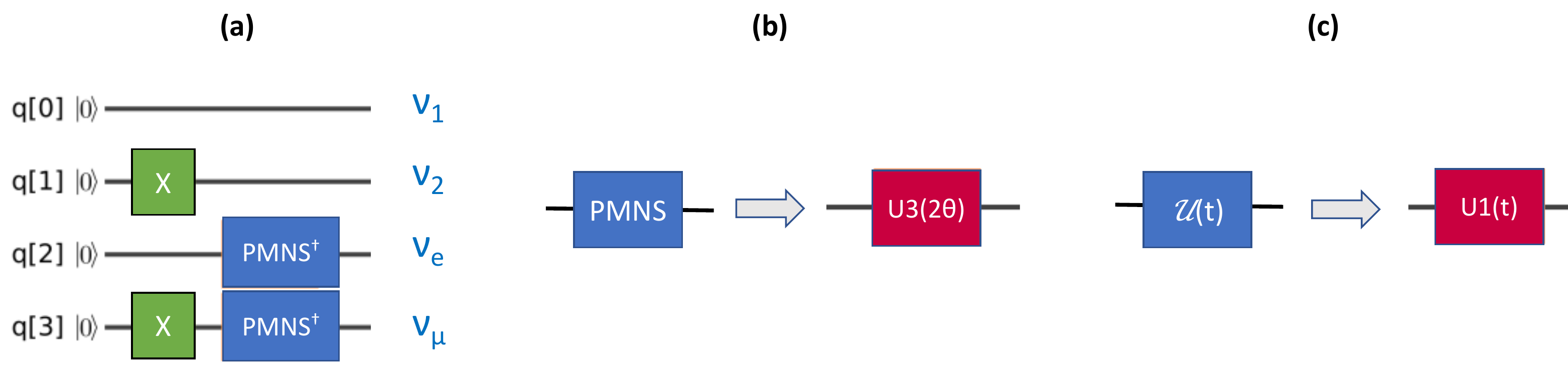}

\includegraphics[width=1.7\columnwidth]{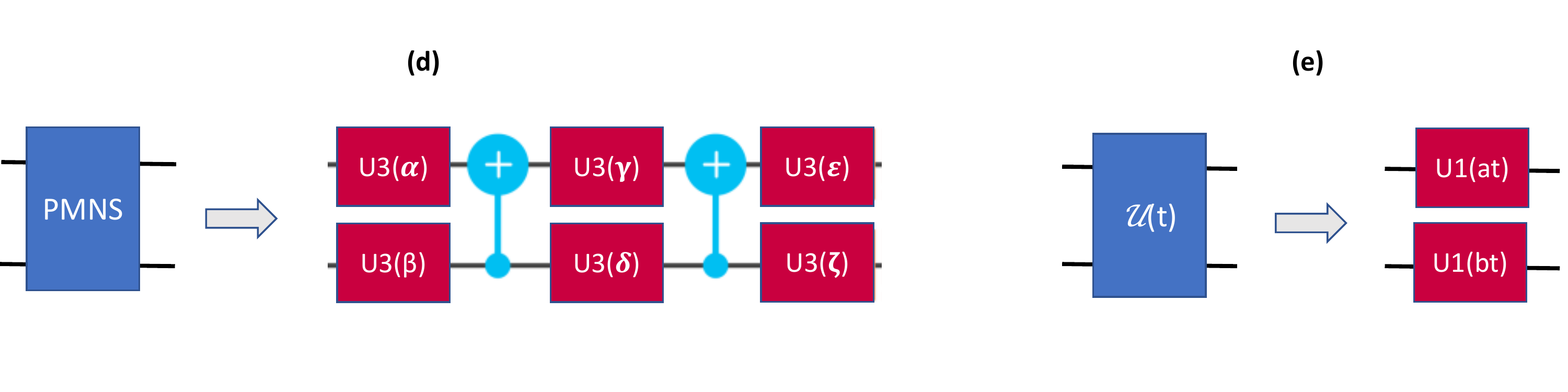}
\par\end{centering}
\caption{Top: Preparation of neutrino flavor and mass basis states in the 2$\times$2 quantum
computation (a), and PMNS (b) and time evolution (c) gates; Bottom: PMNS (d) and time evolution (e) gates in the 3$\times$3 quantum computation. \label{fig:Preparation-of-neutrino}}
\end{figure*}

We thus connect to the conventionally defined $2\times2$ neutrino
PMNS matrix via the definition
\begin{equation}
U_{\rm PMNS}^{2x2}=U3(2\theta,0,0)=\left(\begin{array}{cc}
\cos\theta & -\sin\theta\\
\sin\theta & \cos\theta
\end{array}\right).
\label{eq:pmsn_2x2}
\end{equation}

The IBM-Q $U3$ gate used in this way has a simple interpretation, as a rotation around the Y axis in the Pauli representation. To prepare a neutrino flavor state, we can apply the PMNS operation either to
the $|0\rangle$ state to prepare a $|\nu_{e}\rangle$, or to the
$|1\rangle$ state to prepare a $|\nu_{\mu}\rangle$. The input qubits
in a quantum computation conventionally initialize to $|0\rangle$, and 
the $|1\rangle$ state can be prepared by application of the Pauli-X gate,
$|1\rangle=X|0\rangle$. The preparation of electron and muon neutrino
flavor states, as well as $m_{1}$ and $m_{2}$ mass states in the two-flavor basis
is shown in terms of quantum circuit elements in Fig.~\ref{fig:Preparation-of-neutrino}.

\begin{figure}[b!]
\begin{centering}
\includegraphics[width=0.99\columnwidth]{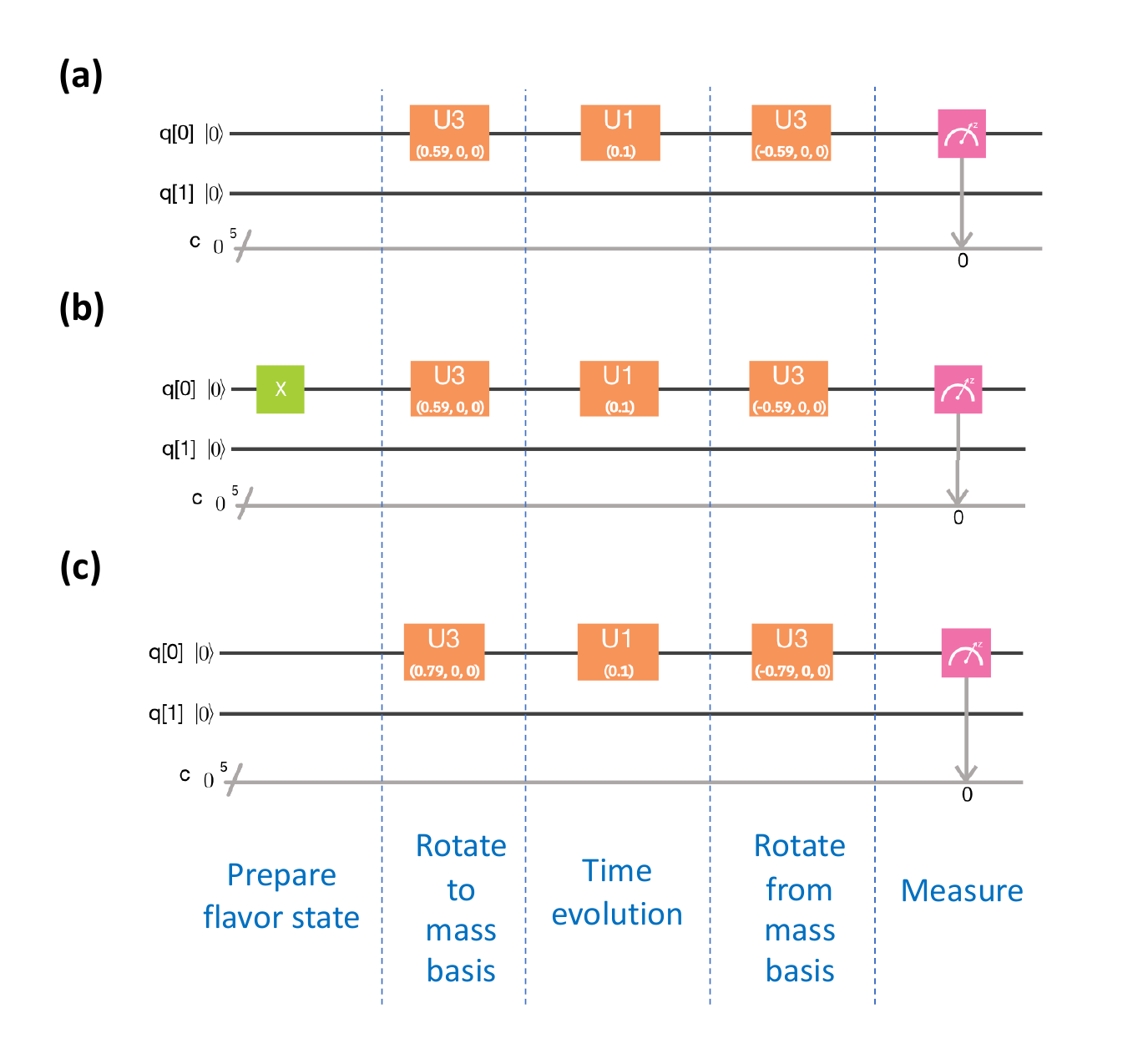}
\par\end{centering}
\caption{Quantum circuits embodying two-flavor neutrino oscillation for (a) $\nu_e$ disappearance; (b) $\nu_\mu\rightarrow\nu_e$; (c) $\nu_\mu\rightarrow\nu_\tau$. \label{fig:Quantum-circuit-embodying}}
\end{figure}

\begin{figure}[h]
\begin{centering}
\includegraphics[width=0.9\columnwidth]{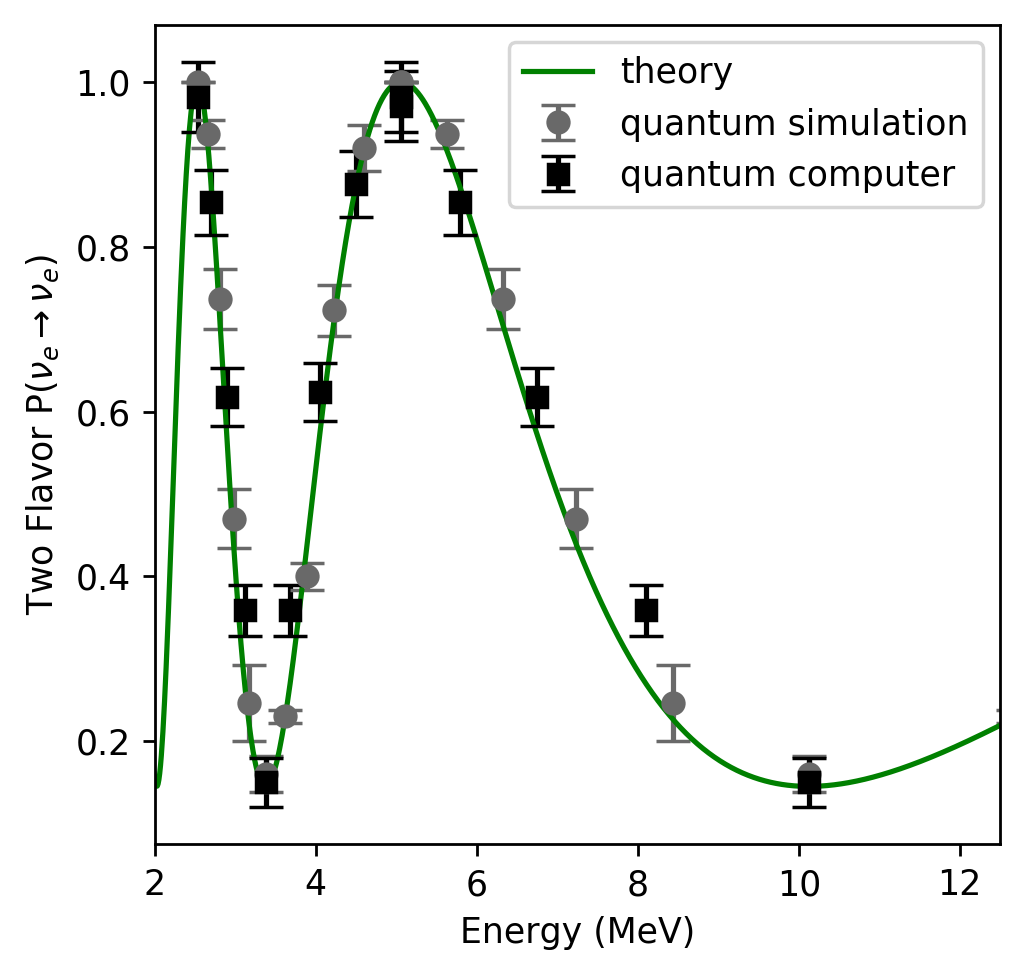}
\par\end{centering}
\caption{Two-flavor electron-neutrino survival probability as a function of the neutrino energy. The green line shows the theoretical calculation using a classical computer. The black circle markers indicate a quantum computer simulation and the black square markers are the output of the IBM Q quantum computer.\label{fig:TwoFlavorCalc}}
\end{figure}

Oscillation probabilities can be calculated by time-evolving the initial flavor state vector in with the appropriate time-evolution operator $\cal{U}$ and then measuring in the flavor basis.  Only relative phases between mass states are relevant for oscillations, and so without
loss of generality, we can measure all phases relative to the $m_{1}$
basis state. The time-evolution
operation can thus be encoded in an $S$ gate:
\begin{equation}
{\cal U}(t)=S(\phi)=\left(\begin{array}{cc}
1 & 0\\
0 & e^{i\phi}
\end{array}\right),\label{eq:SGate}
\end{equation}
where $\phi=\Delta m^{2}t/2E\hbar$. In the two-flavor system, we thus
find a particularly simple representation of the PMNS and time-evolution ``gates'',
 shown in Fig.~\ref{fig:Preparation-of-neutrino} center and right. Examples
of circuits that realize various two-flavor oscillation scenarios
are given in Fig.~\ref{fig:Quantum-circuit-embodying}. With the quantum circuit defined, we can proceed to evaluate oscillation probabilities on the quantum processor. We run 1024 trials and count flavor measurement outcomes to establish oscillation probabilities in the two-flavor system.  Fig.~\ref{fig:TwoFlavorCalc}, shows the comparison of
the quantum calculation to the theoretical expectation for parameters
relevant to $\nu_e$ disappearance at the KamLAND experiment~\cite{Eguchi:2002dm} as an example. The figure compares actual quantum
computations, calculated on the IBM quantum processor (squares), simulated runs of the quantum computer, which represent the same operations performed without decoherence or errors (dots), and the standard two-flavor oscillation formula (lines). The quantum evolution
matches very well with expectations from both theory and quantum simulation.

For this circuit and others described in this paper we have chosen gate arrangements that are simple and intuitive, mirroring the quantum operations involved in physical neutrino oscillations.  It is, however, clear that this physically motivated sequence is not necessarily the most efficient way of performing the relevant unitary operation on a quantum computer.  The three gates of Fig.~\ref{fig:Preparation-of-neutrino} that represent flavor-rotation, time-evolution, and inverse flavor rotation could, for example, be combined into a single U(3) gate. Similar simplifications are possible for the other circuits presented in this paper.  Since our goal in this work is to illustrate the physical encoding of the neutrino oscillations system into a quantum computer, we have opted for the more intuitive, physically motivated circuit layouts throughout.

\section{Three-flavor neutrino oscillation\label{sec:ThreeNu}}

A three-flavor neutrino oscillation involves a Hilbert space of dimension
three, requiring more than one qubit. The minimal representation
can be encoded on two qubits, via a basis definition such as:
\begin{equation}
\begin{array}{c}
|00\rangle\rightarrow|\nu_{1}\rangle=(1,0,0,0),\quad|01\rangle\rightarrow|\nu_{2}\rangle=(0,1,0,0)
\\
|10\rangle\rightarrow|\nu_{3}\rangle=(0,0,1,0),\quad|11\rangle\rightarrow|\nu_{X}\rangle=(0,0,0,1)
\end{array}
\end{equation}
There is one redundant basis state, $|\nu_{X}\rangle$, in this representation.
This could represent a fourth neutrino flavor in models with sterile neutrinos, but for the present example we will consider it as physically decoupled, and thus unphysical. As in the two-flavor case, to prepare a flavor state, we must apply the PMNS operation to an initial state in the computational basis. Unlike in the two-neutrino example, however, creating a set of quantum gates to implement the PMNS operation on two entangled qubits is non-trivial. A real unitary two-qubit gate requires at least two CNOT and 12 elementary gates~\cite{vatan2004optimal} for an entirely general representation. Constraints on the PMNS matrix due to re-phasing invariance may be expected to allow for a more compact representation. Following exploration of several possibilities, we constructed a parameterizable set of six real U3 gates acting
on two qubits $A$ and $B$, with two interspersed CNOT gates to reproduce the PMNS operation.  To fix the free parameters of this arrangement, we map the circuit onto matrix multiplication in the computational basis, and perform a numerical fit to match its entries to the experimentally determined PMNS matrix elements~\cite{Esteban:2018azc}. We fit for the PMNS matrix and invert the gate arrangement for PMNS$^\dagger$. The PMNS and PMNS$^\dagger$ operations are thus constructed as:
\begin{eqnarray}
\begin{array}{c}{\rm PMNS}=U3_{A}(\epsilon)\,U3_{B}(\zeta)\,CNOT_{AB}\quad\times\quad\quad\\ \quad\quad\,U3_{A}(\gamma)\,U3(\delta)_{B}\,CNOT_{AB}U3_{A}(\alpha)\,U3(\beta)_{B},\end{array}\\
\begin{array}{c}
{\rm PMNS}^{\dagger}=U3_{A}(-\alpha)\,U3_{B}(-\beta)\,CNOT_{AB}\quad\times\quad\quad\\
\quad\quad\,U3_{A}(-\gamma)\,U3_{B}(-\delta)\,CNOT_{AB}U3_{A}(-\epsilon)\,U3(-\zeta)_{B}.
\end{array}
\end{eqnarray}
The best-fit parameters $\alpha$,$\beta$,$\gamma$,$\delta$,$\epsilon$, and $\zeta$ (primed and unprimed) reproduce the measured PMNS and PMNS$^\dagger$ elements within one part in 10$^6$ when no CP violating phase is present-- comfortably within experimental uncertainty. This parameterization can be extended with two additional U3 gates and a CNOT in order to incorporate a Dirac CP phase with similar accuracy. These parameters are tabulated in Appendix B. The PMNS gate decomposition in terms of component gates is shown diagrammatically in the bottom left panel of Fig.~\ref{fig:Preparation-of-neutrino}.
\begin{figure}[b!]
\begin{centering}
  \includegraphics[width=0.9\columnwidth]{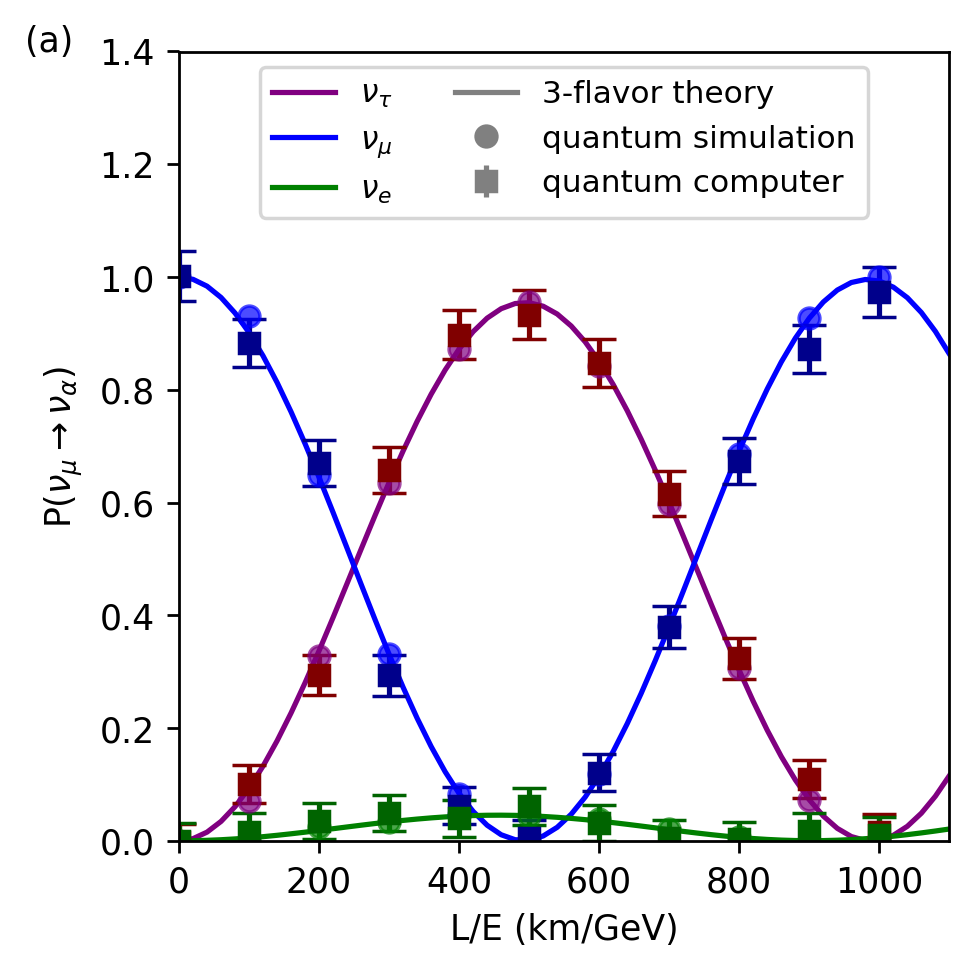}\\
  \includegraphics[width=0.9\columnwidth]{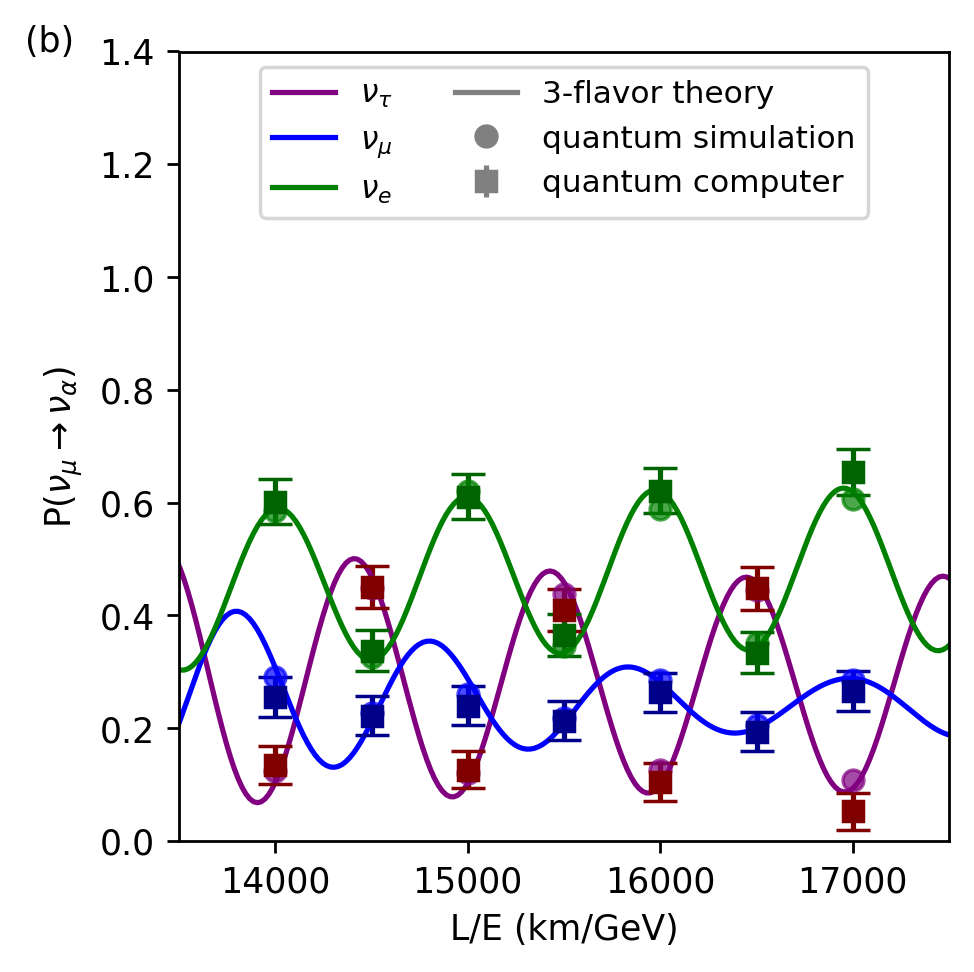}
\par\end{centering}

\caption{Calculations of three-flavor neutrino oscillations evaluated using
a quantum computer (squares), quantum computer simulator (circles),
and theory (lines). The quantum computer results have been corrected
for gate read errors based on the matrix $M$ described in the text. The two plots (a) and (b) show three-flavor oscillation probabilities calculated in two characteristic $L/E$ ranges.\label{fig:Calculations-of-three-flavor}}
\end{figure}

Once the initial flavor state is prepared, the time-evolution operation $\cal{U}$ must be
implemented. This is represented in the computational
basis by:
\begin{equation}
\mathcal{U}(t)=\exp\left[i\,{\rm diag}\left(0,\, \Delta m_{12}^{2}\frac{t}{2E\hbar},\, \Delta m_{13}^{2}\frac{t}{2E\hbar},\, \Phi \right)\right],
\end{equation}where $\Phi$ is an arbitrary phase that can be picked for convenience, since the fourth basis
state is unobservable. A straightforward choice that can be implemented as one-qubit gates acting on $A$ and $B$ is:
\begin{equation}
{\cal U}(t)=S_{A}\left(i\Delta m_{13}^{2}\frac{t}{2E\hbar}\right)S_{B}\left(i\Delta m_{12}^{2}\frac{t}{2E\hbar}\right);
\end{equation}
where $S_A$ and $S_B$ are the $S$ gates for qubits $A$ and $B$, respectively.

The complete quantum circuit for the oscillation calculation in the three-neutrino space comprises of state preparation, time-evolution and flavor measurement, and is shown in Appendix A, Fig.~\ref{fig:Circuit}. This circuit can be run repeatedly to prepare qubits in flavor eigenstates, time-evolve them, and measure their flavor after propagation in order to establish oscillation probabilities.
Since this is a substantially more complex circuit than the two-flavor case, gate errors and read errors are expected to be more prevalent. We correct for the effects of read errors in the final oscillation probability by applying an inverted error matrix in the computational basis, which accounts for decoherence and read errors in a statistical manner, based on qubit readout accuracy measured using runs with $L/E=0$. More details on this procedure can be found in Appendix C.

Figure~\ref{fig:Calculations-of-three-flavor}
shows two example calculations of three-flavor oscillation probabilities given an initial muon neutrino beam. The first panel shows calculations at smaller
$L/E$ where the oscillation is effectively a two-flavor system. The second panel shows the behavior near the first oscillation maximum where three flavors $\nu_{e},\nu_{\mu},\nu_{\tau}$, are participating strongly. Good agreement with theory is observed in both regimes. In both cases, the electron
flavor is slightly over-represented, potentially due to read and gate
errors that are not entirely symmetrically distributed between flavors. The size of the effect is comparable to the statistical and systematic uncertainty, which receives contributions from: 1) accumulated gate errors,
based on the {\tt ibmqx2} spec of $\sim10^{-3}$ per gate added in quadrature
over 50 gates and 2) statistical uncertainty, from the finite number
of evolutions (1024) used to establish the oscillation probability. After running the simulation and applying a statistical read out error mitigation tuned on zero-time simulations (explained in Appendix
C), strong agreement between the quantum and classical computations are obtained.

\section{Neutrino oscillations with new physics}

In addition to standard neutrino oscillations, BSM effects that have been searched for in neutrino oscillation experiments can be incorporated into the quantum circuit straightforwardly, by either a) extending the PMNS matrix and time-evolution operator to higher dimensionality or b) introducing new effects in the time-evolution term. Here we briefly review a few of these scenarios.

\subsection*{Sterile neutrinos}

As we have seen, the incorporation of at least one additional basis state within the Hilbert space is mandatory, given a two-qubit realization.  To use this state to represent an oscillating forth neutrino, as suggested by short baseline neutrino anomalies~\cite{Aguilar-Arevalo:2018gpe,Kopp:2013vaa,Mention:2011rk,Aguilar:2001ty}, two adjustments are required: 1) extension of the PMNS matrix to mixing in four dimensions, which is already achievable in our present parametrization, given appropriate gate coefficients; and 2) independent control of the phase of the $|\nu_4\rangle=|11\rangle$ mass state in the time evolution operator. This is necessarily an operation that involves entangling the two qubits, and so cannot be implemented on single-qubit gates only.  A circuit that produces the required time evolution (an independently specified phase on each of $|01\rangle$, $|10\rangle$, and $|11\rangle$) is shown in Fig.~\ref{fig:BSMCircuits}, top.  This circuit configured with parameters:
\begin{eqnarray}
\phi_{1}=\frac{1}{4E\hbar}\left(\Delta m_{12}^{2}-\Delta m_{13}^{2}+\Delta m_{14}^{2}\right),\\
\phi_{2}=\frac{1}{4E\hbar}\left(\Delta m_{12}^{2}+\Delta m_{13}^{2}-\Delta m_{14}^{2}\right),\\
\phi_{3}=\frac{1}{4E\hbar}\left(-\Delta m_{12}^{2}+\Delta m_{13}^{2}+\Delta m_{14}^{2}\right)
\end{eqnarray}
will achieve the necessary four-state time evolution needed to implement quantum simulations of the four-flavor neutrino system extended for a single sterile neutrino.

\subsection*{Non-standard interactions and matter effects}
The modelling of either standard~\cite{Mikheev:1986gs,Wolfenstein:1977ue} or non-standard~\cite{Antusch:2008tz,Davidson:2003ha} matter effects, with or without violations of Lorentz symmetry, can be incorporated without changing the three-flavor oscillation quantum circuit by adjustment of the input parameters that describe the PMNS and time-evolution operations in the modified matter basis.  A discussion of the parametrizations that incorporate these effects is given in Appendices D and E.

\subsection*{Decoherence}
\begin{figure}
\begin{centering}
\includegraphics[width=0.99\columnwidth]{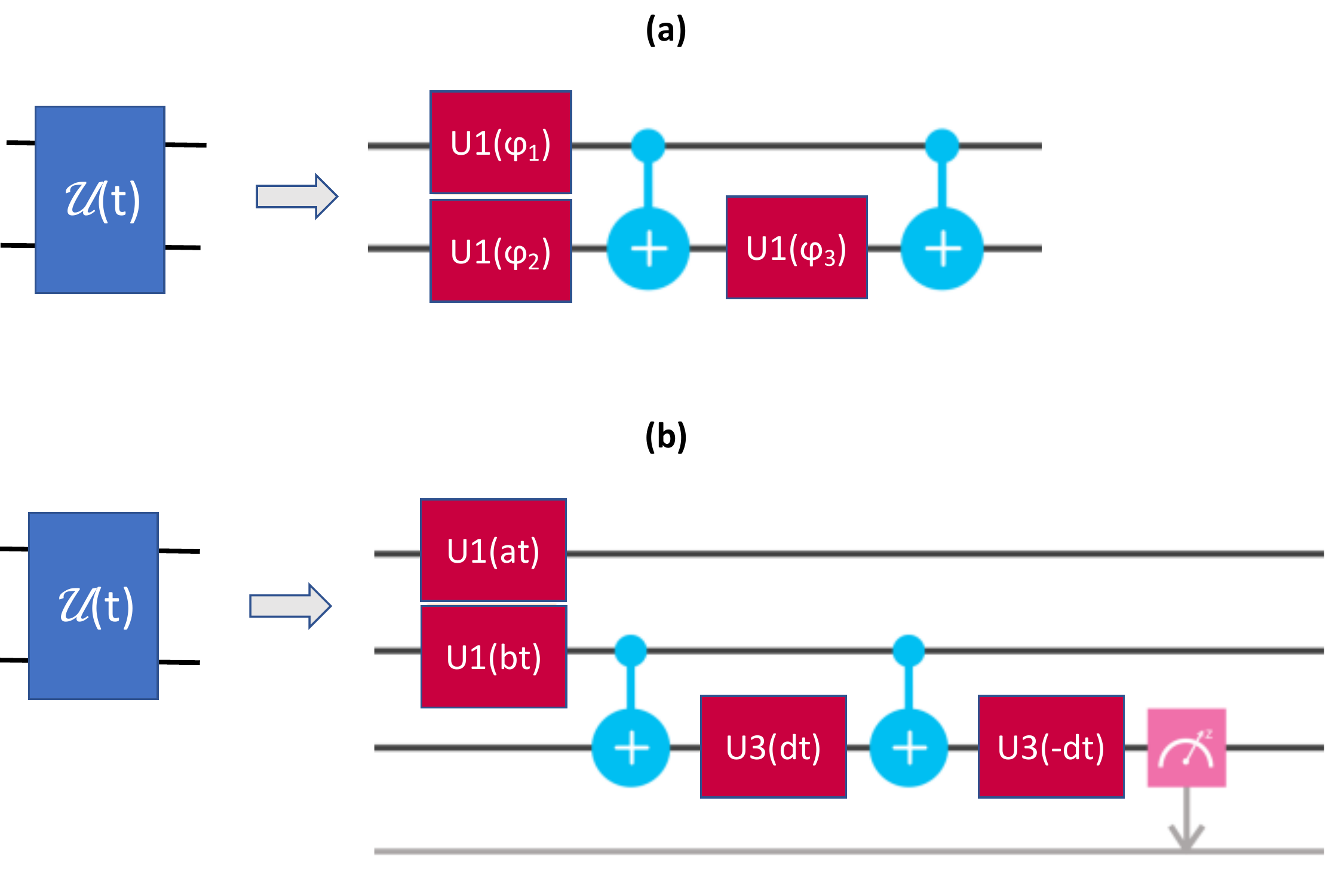}
\par\end{centering}

\caption{Quantum circuits for two BSM oscillation scenarios. (a) sterile neutrino oscillations; (b) anomalous decoherence in the mass basis.\label{fig:BSMCircuits}}
\end{figure}

Decoherence is a non-standard neutrino oscillation effect~\cite{Lisi:2000zt,Carpio:2017nui,Gago:2000qc} that is often considered in connection with quantum gravity or spacetime foam models~\cite{Brustein:2001ik}.  In decoherence scenarios, development of entanglements between parts of the neutrino wave function and an external environment lead to partial collapse of the wave function and suppression of oscillations.  Decoherence can be manifest in various bases, depending on the degrees of freedom within the neutrino subsystem that the environment entangles. Fig.~\ref{fig:BSMCircuits}, bottom, illustrates a quantum circuit that implements decoherence in neutrino oscillations via generation of entanglements in the mass basis for small $dt$. In this circuit, an auxiliary qubit representing the environment is initialized to zero and acquires a small and time-dependent admixture of the $|1\rangle$ basis state, if and only if the second neutrino qubit is in a $|1\rangle$ state.  This entanglement between the system and the ancilla qubit acts to suppress coherence in the off-diagonal elements of the neutrino system density matrix.  The process of continuous measurement during evolution for large $dt$ is approximated by repeated units of time evolution, entanglement and measurement, resetting the ancilla to zero after each time step.  In such a scheme, entanglement is developed between the system and ancilla, suppressing the off-diagonal system density matrix elements.  This entanglement is then conveyed to the outside world via measurement. The process of measurement does not itself generate decoherence, but allows the ancilla to be disentangled and reset to zero for the next block of time evolution and further suppression of the off-diagonal system density matrix elements.  Such a scheme could model gravitational decoherence, or flavour change through wave packet separation given a normal ordering of neutrino masses (with one mass state much heavier than the others). 

\section{Conclusions}

We have demonstrated a quantum mechanical simulation of neutrino oscillations on a quantum computer, using both two-flavor and three-flavor systems.  The two-flavor system has an almost trivial realization in the two-dimensional Hilbert space of a single qubit, with implementation of PMNS and time evolution gates using individual single-qubit gates.  The three neutrino system, on the other hand, requires a more complicated quantum circuit, involving the entanglement of two qubits to produce a Hilbert space of four dimensions.  A subspace of three of these states is used for the calculation.  Our implementation of the 3$\times$3 PMNS matrix on the two qubit space in terms of six general rotation gates and two CNOTs provides an accurate and economical realization of the PMNS operation within a quantum circuit.  By choosing a phase convention such that phases are measured relative to $\nu_1$ and allowing for phase freedom in the unphysical fourth basis state, the time-evolution operator can be implemented using two single-qubit gates.

Quantum calculations using both the two- and three-flavor system agree with theoretical expectations for the neutrino oscillation probability within systematic and statistical uncertainty.  Although in agreement, the calculation presented here is characteristically different to the classical computation of oscillation probabilities, since the qubits act as direct quantum analogues to the evolving neutrino flavor wave function.  The system is prepared coherently, evolves forward in time unitarily, and has its wave function collapsed to measure the final flavor oscillations, just as in a neutrino oscillation experiment.   

Analogues of real quantum systems inside quantum processors such as the one presented in this paper may eventually enable computations that surpass the capabilities of their classical counterparts.  This is especially likely for strongly coupled or highly entangled systems, such as collectively oscillating neutrinos in supernovae, for example. Understanding how to translate simple and well-understood calculations into quantum circuits is a necessary first step toward realizing this goal.  In this work, we have presented one such example, creating an analogue to two and three-flavor neutrino oscillations inside a publicly accessible quantum processor.  This may be the first of many neutrino oscillation calculations to profit from the power of quantum information processing technologies.

\section*{Acknowledgements}
We thank Jean DeMerit, Kareem R. H. A. M. Farrag, Roxanne Guenette, Jonathan Asaadi  and Peter Denton for proof-reading this manuscript and providing insightful comments.

BJPJ is supported by the Department of Energy under award numbers DE-SC0019054 and DE-SC0019223.
CAA is supported by U.S. National Science Foundation (NSF) grant No. PHY-1801996. BJPJ and CAA thank the organizers of the workshop ``New Opportunities at the Next Generation Neutrino Experiments'' at the University of Texas at Arlington, where this work was completed.
We thank IBM for making the public quantum processor available to the world- an invaluable resource as we prepare to move towards the quantum revolution.  We acknowledge use of the IBM Q for this work. The views expressed are those of the authors and do not reflect the official policy or position of IBM or the IBM Q team.

\bibliography{main}

\clearpage
\newpage
\appendix

\section{Running quantum computation on {\tt IBM-Q}} 
Computations were run on two IBM publicly accessible quantum computers {\tt ibmqx2} (Yorktown).  The two least read-error-prone qubits on this five qubit machine that could be connected by the appropriate logic gates within the allowable topology were chosen to support the computational basis. At the time of writing these were qubits 0 and 2, with reported gate errors of 0.77$\times 10^{-3}$ and 1.03 $\times 10^{-3}$, and read errors of $7.6\%$ and $2.9\%$ respectively, and a mulit-qubit read error of 2.21$\%$.   Some supplementary calculations were also run on IBM-Q {\tt ourense}.

It is difficult to convert such gate-wise error specifications into an expected calculational accuracy, so we instead opted to measure the accuracy directly by running $L/E=0$ simulations. Some contribution to the error budget is expected from errors that accumulate only when the time evolves a finite amount, so we conservatively associate an additional contribution for gate error on top of that measured from unoscillated points.  The complete quantum circuit as implemented on {\tt ibmqx2} is shown in Supplementary Fig.~\ref{fig:Circuit}.  This was run with 1024 shots to establish the survival probability, and the statistical error associated with this count is included in our uncertainty budget.

As well as access to the quantum processor, the IBM website offers a quantum simulator tool to simulate the circuit before running it.  In all cases, simulations agreed near perfectly with theoretical expectations.  This suggests that wherever the data show small differences from theoretical expectations, these are to be attributed to the imperfections of the quantum processor.

\begin{figure*}
\begin{centering}
\includegraphics[width=1.8\columnwidth]{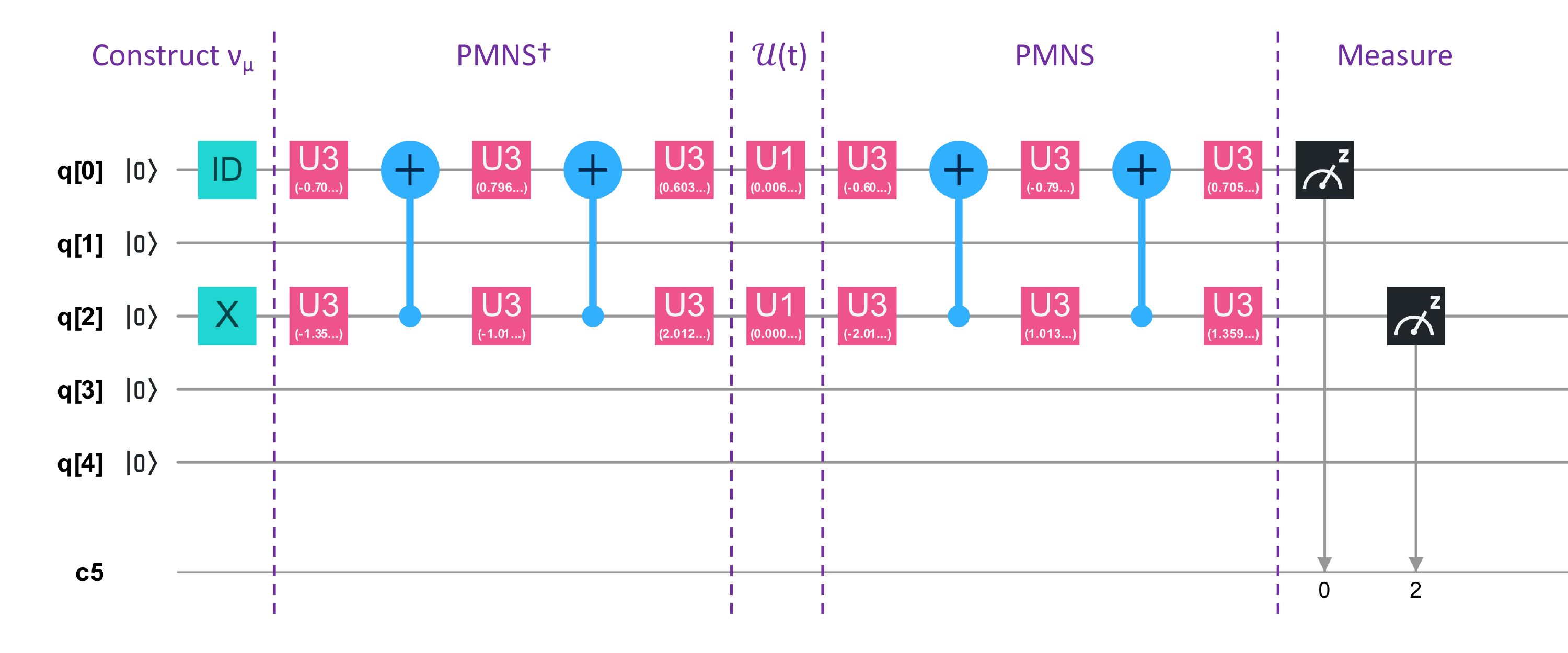}
\par\end{centering}
\caption{Three-flavour neutrino oscillation experiment as run on the IBM quantum computer.\label{fig:Circuit}}
\end{figure*}

\section{Fitting the PMNS matrix gate parameters}
The fit to the PMNS matrix is made by gradient minimization over six parameters. We minimize the sum of squared residuals of each of the elements of the matrix. The final fit converges to the true PMNS matrix within one part in 10$^6$ for every element, comfortably within experimental uncertainty.  The input values of the PMNS matrix used in the fit are from~\cite{Esteban:2018azc} and are:
\begin{equation}
U_{PMNS}=\left(\begin{array}{cccc}
0.821427 & 0.550313 & 0.149708 & 0\\
-0.481513 & 0.528538 & 0.699138 & 0\\
0.305618 & -0.646377 & 0.699138 & 0\\
0 & 0 & 0 & 1
\end{array}\right)
\end{equation}
The bottom row represents a decoupled, unphysical state, but is necessary to span the four-dimensional Hilbert space represented by two qubits.  The best-fit parameters from fitting the PMNS and PMNS$^\dagger$ matrices in the computational basis are:
\begin{equation}
\begin{array}{c}\alpha=-0.6031\quad\beta=-2.0125\quad\gamma=0.7966\\\delta=1.0139\quad\epsilon=0.7053\quad\zeta=1.3599\\
\end{array}
\end{equation}
This gate configuration will apply the PMNS rotation to any state prepared in the mass basis, as defined in this work.

\begin{widetext}
\begin{equation*}
M=\left(\begin{array}{cccc}
\left(1-f_{1}\right)\left(1-f_{2}\right)+f_{1}^{2}+f_{2}^{2} & (1-f_{1})f_{2}-f_{2}^{2} & (1-f_{2})f_{1}-f_{1}^{2} & f_{1}f_{2}\\
(1-f_{1})f_{2}-f_{2}^{2} & \left(1-f_{1}\right)\left(1-f_{2}\right)+f_{1}^{2}+f_{2}^{2} & f_{1}f_{2} & (1-f_{2})f_{1}-f_{1}^{2}\\
(1-f_{2})f_{1}-f_{1}^{2} & f_{1}f_{2} & \left(1-f_{1}\right)\left(1-f_{2}\right)+f_{1}^{2}+f_{2}^{2} & (1-f_{1})f_{2}-f_{2}^{2}\\
f_{1}f_{2} & (1-f_{2})f_{1}-f_{1}^{2} & (1-f_{1})f_{2}-f_{2}^{2} & \left(1-f_{1}\right)\left(1-f_{2}\right)+f_{1}^{2}+f_{2}^{2}
\end{array}\right)\label{eq:M}
\end{equation*}
\end{widetext}

\section{Correcting for bit flips and decoherence}

Some random qubit readout errors are naturally expected for any quantum computation.  However, we can correct our final distributions against some of these information losses statistically.  We consider that there is some average probability $f_1$ and $f_2$ for either of the qubits to be flipped leading to an incorrect flavor measurement, which is approximately uniform across circuits.  Then, the effect on the final distribution of events in ($\nu_e$,$\nu_\mu$,$\nu_\tau$,$\nu_X$) space is to multiply final state distributions by a matrix M, given in Eq.~\ref{eq:M}.

The form of $M$ can be understood by considering that spurious transitions $\nu_{\mu}\leftrightarrow\nu_{e}$,
$\nu_{\tau}\leftrightarrow\nu_{e}$, $\nu_{\tau}\leftrightarrow\nu_{s}$
and $\nu_{\mu}\leftrightarrow\nu_{s}$ require only one qubit flip in the final read,
whereas spurious transitions $\nu_{e}\leftrightarrow\nu_{s}$, $\nu_{\mu}\leftrightarrow\nu_{\tau}$
require two.  $f_1$ and $f_2$, the rates of bit flips in qubits 1 and 2, can be measured by examining the rate of spurious transitions at $L/E=0$ where no physical oscillation effects are expected, and any transformation must be associated to spurious bit flips. The quantum simulator predicts zero transformation at $L/E=0$, whereas we find the quantum computer gives random some transformations, consistent with $f_1\sim13$\% and $f_2\sim3$\%. With $f$ measured, we can invert $M$ and apply this inverted matrix to correct the final probability distributions.  This correction is applied in order to obtain our final data / theory comparisons.

\section{Incorporation of matter potentials and non-standard interactions}

When neutrinos travel through a medium they experience a potential produced by coherent-forward scattering with electrons, protons, and neutrons. The potential sourced by protons and neutrons is the same for all neutrino flavors, and thus induces an overall phase in the neutrino system evolution. The electron charged-current potential is flavor asymmetric, producing an observable modification in the neutrino oscillation probability. The matter potential can be written in the flavor basis as 
\begin{equation}
    V_m = \sqrt{2} G_F 
\left( 
\begin{array}{ccc}
   N_e & 0 & 0 \\
   0 & 0 & 0 \\
   0 & 0 & 0
\end{array} 
\right),
\label{eq:matter_potential}
\end{equation}
where $G_F$ is the Fermi constant and $N_e$ is the electron number density. Then the total neutrino Hamiltonian can be written as 
\begin{equation}
H = H_{vac} + V_m = U_m^\dagger \Delta U_m.
\label{eq:hamiltonian_decomposition}
\end{equation}
The Hamiltonian can be diagonalized by a unitary transformation, $U_m$, that relates the flavor basis to the Hamiltonian eigenstates, and a diagonal matrix, $\Delta$, which contains the energy eigenvalues. In a two flavor system, $U_m$ can be written as a $2\times 2$ rotation matrix analogous to vacuum case~\ref{eq:pmsn_2x2}, where the rotation angle is given by
\begin{equation}
    \theta_m = \arctan{\frac{\Delta\sin^2 2\theta}{\Delta \cos 2\theta_0 - \sqrt{2} G_F N_e}},
\label{eq:matter_angle}
\end{equation}
where $\Delta = \Delta m^2/(2E)$ and $\theta_0$ is the vacuum mixing angle. The relevant Hamiltonian eigenvalue difference is given by
\begin{equation}
  \lambda = \sqrt{\left(\Delta \cos 2 \theta_0 - \sqrt{2} G_F N_e \right)^2 + \Delta^2 \sin^2 2 \theta_0}.
\label{eq:matter_eigenvalue}
\end{equation}
Thus the matter modification does not require a new quantum circuit; we can simply replace the vacuum mixing angle and eigenvalue for the expressions given above. For the case of the three-neutrino scenario exact expressions for the effective mixing angles and eigenvalues are lengthy, but can be readily found by numerical diagonalization.

Effects of the standard neutrino matter potential have been observed with natural sources, {\it e.g.} in Solar neutrino experiments, and with human-made sources, {\it e.g.} in accelerator neutrinos experiments. Deviations from the standard potential can be due to new forces that manifest themselves as vector or scalar interactions. These have been constrained by searches of anomalous neutrino flavor changing and, more recently, in coherent-scattering neutrino experiments. The status of recent constraint on non-standard interactions can be found in Ref.~\cite{Esteban:2018ppq}, where the constraints are given relative to the weak-force strength, $G_F$, and flavor dependent coefficients, $\epsilon_{\alpha \beta}$. Depending on the target and the flavor structure these are constrained from O(1\%) to O(10\%). The effects of vector non-standard interactions can also be calculated using a quantum processor by using PMNS gate with effective matter potential mixing angles. The appropriate angles and eigenvalues can be determined by diagonalizing the following Hamiltonian in Eq~\ref{eq:hamiltonian_decomposition} where one ought to replace $V_m$ by
\begin{equation}
    V_m^{nsi} = \sqrt{2} G_F N_e 
\left( 
\begin{array}{ccc}
   1 + \epsilon_{ee} & 0 & 0 \\
   0 & 0 & 0 \\
   0 & 0 & 0
\end{array} 
\right).
\label{eq:nsi_matter_potential}
\end{equation}

\section{Incorporation of Lorentz Violation}

The Standard Model (SM) of particle physics~\cite{Weinberg:1967tq} can be thought as an effective field theory towards a grand unified theory of nature~\cite{Weinberg}. In many extensions of the SM, Lorentz symmetry is broken. Neutrinos, as natural interferometers, are extremely sensitive to high-scales where Lorentz Violation (LV) may be manifest. In fact, neutrinos have some of the strongest constraints on LV non-renormalizable operators~\cite{Aartsen:2017ibm}. Calculations of oscillation probabilities in the presence of LV can be performed using the quantum circuits presented in this paper. LV can be incorporated in the neutrino Hamiltonian in the following way:
\begin{equation}
    H = H_{vac} + \tilde a^{3} + \tilde c^{4} E + \tilde a^{5} E^2 + \tilde c^{6} E^3 + ...,
\end{equation}
where $\tilde a^{d}$ ($\tilde c^{d}$) is a matrix that contains the strength of interaction between the neutrino and a Lorentz violating field produced by a CPT even (odd) effective operator of dimension $d$. Similarly to the case of matter interactions, this Hamiltonian can be diagonalized numerically in order to obtain appropriate effective mixing angles and frequencies. One can then use the PMNS quantum gate and evolution operators discussed in the main text.

\end{document}